\documentstyle[preprint,aps]{revtex}

\begin{document}
\draft
\preprint{ }

\title{Participation-Ratio Entropy and Critical Fluctuations in
the Thermodynamics of Pancake Vortices}

\author{Jun Hu$^{1,2}$ and A.H. MacDonald$^{1}$}

\address{$^1$Department of Physics, Indiana University,
 Bloomington, Indiana 47405}

\address{$^2$National High Magnetic Field Laboratory,
 Florida State University, Tallahassee, FL 32306}

\date{\today}
\maketitle

\begin{abstract}

We report on a study of the thermodynamics of the Ginzburg-Landau model for
two-dimensional type-II superconductors near
the transition from the normal state to the Abrikosov-lattice state.
We couch our analysis in terms of the participation-ratio
entropy, $s(P)$, which expresses the volume in
order-parameter-space with a given participation-ratio
for the local superfluid density.
$s(P)$ completely determines the thermodynamics of the system.
We report on results for $s(P)$ obtained analytically using
perturbation expansion methods and numerically using Monte Carlo
simulation methods and discuss the weak first-order phase transition which
occurs in this system in terms of the properties of
$s(P)$.

\end{abstract}


The discovery of high-temperature superconductivity in materials
with strong planar anisotropy has led to renewed
experimental\cite{beck1,urbach1,safar1,kwok1}
and theoretical\cite{doniach1,brezin1,fisher5,tesanovic1,huse1}
interest in the properties of two-dimensional
and strongly anisotropic three-dimensional type-II superconductors at fields
near $H_{c2}(T)$ where thermal fluctuations are important.
Phase transitions of type-II superconductors in a magnetic field
are unusual because of the Landau level
degeneracy of Cooper-pair states in a magnetic field\cite{macdonald1}.
In mean-field-theory continuous phase transitions occur simultaneously in
a number of channels equal to the Landau level degeneracy and the
low-temperature Abrikosov lattice state has both
superconducting
and positional order.  The Cooper-pair Landau level degeneracy
also increases the importance of fluctuations and as a result
the phase transition is considerably rounded.
It is well established experimentally and theoretically that the
Abrikosov vortex lattice state in $D = 3$ dimension melts
into a vortex liquid state through a weakly first order
phase transition,~\cite{safar1,kwok1,brezin1,hetzel1,hu3}
A consensus\cite{tesanovic1,kato12,hu1,franz1}
has emerged from recent work that a weak first-order
phase transition with a latent heat which is $\sim 2\%$ of the
mean-field condensation energy at the depressed transition temperature
also survives in $D=2$, although this view is not
universally held\cite{oneil12}.  We have previously\cite{hu1}
introduced a formulation of the thermodynamics of this system
in terms of a quantity, $s(P)$,  which we refer to here
as the participation-ratio
entropy and which measures the volume in order-parameter space associated
with a given participation ratio for the local superfluid density.
In this Letter we report on evaluations of $s(P)$ based on
high-temperature and low-temperature expansions and on
Monte-Carlo simulations of the Ginzburg-Landau (GL) model.

For fields sufficiently close to $H_{c2}$ the order parameter
is confined to the lowest Cooper-pair Landau-level\cite{lllvalidity}.
The free energy density of the lowest-Landau-level GL (LLL-GL)
model is
\begin{equation}
f[\Psi]  = (\alpha (T) + \hbar e H / m^{*} c ) \vert \Psi \vert^2 +
\frac{\beta}{2} \vert \Psi
\vert^4
\label{eq:1}
\end{equation}
where $\alpha(T) = \alpha^{'}(T-T_{c0})$, $T_{c0}$ is the zero-field
transition temperature, $m^{*}$ parameterizes the energy cost of spatial
variation of the order parameter and $\beta$ is taken to be
independent of $T$.  The mean-field theory transition temperature
satisfies
$\alpha_H (T_{c2}) \equiv \alpha (T_{c2}) + \hbar e H / m^{*} c  = 0$.
The LLL-GL model free energy is
$ F_{GL} \equiv \int d \vec r  f [ \Psi ( \vec r)]$.
For $ T < T_{c2}$ the quadratic term in Eq.~(\ref{eq:1}) lowers
$F_{GL}$ while the quartic term always makes a positive contribution.
Our approach to the thermodynamics of this system is based on the
observation that for a given magnitude of the quadratic term in
$F_{GL}[\Psi]$ the quartic term is larger when the order parameter has
a smaller participation ratio\cite{partratio},
\begin{equation}
P[\Psi] = { (\int d^2 \vec r |\Psi |^2 )^2 \over
A \int d^2 \vec r |\Psi|^4 }.
\label{eq:a1}
\end{equation}
$P[\Psi]$ is roughly the fraction of the area $A$ of the sample
over which the local superfluid density ($|\Psi|^2$) is spread.
(In LLL-GL theory $P^{-1}[\Psi] $
is known as the Abrikosov ratio, $\beta_A[\Psi]$.)  For
a given participation ratio the minimum of $F_{GL}/A$
is $ - \alpha_H^2 P /2 \beta $.  Any order parameter
in the LLL has one zero\cite{tesanovic1,arovas1} (vortex) for each
magnetic flux quantum through the system and therefore cannot
have the constant value required to obtain $P=1$.
The mean-field-theory order parameter of the LLL-GL model
is the LLL order parameter with the maximum value of $P$; $P$ is maximized by
placing the vortices on a triangular lattice; $P_{\bigtriangleup}
= 1/\beta_{A\bigtriangleup} = 0.862370 \cdots $.
At finite temperatures
the properties of the LLL-GL model are determined by a competition
between the thermal weighting factor ($ \exp ( - F_{GL}[\Psi] / k_B T) $)
which favors large values of $P[\Psi]$ and the distribution function
of $P[\Psi]$ which is peaked at smaller values of $P$ as we discuss
below.

We choose to work in the Landau gauge ($\vec A = (0, Bx, 0)$).
The order parameter $ \Psi (\vec r) $ can then be
expanded in the form,~\cite{hu1}
\begin{equation}
\Psi (\vec r) =  (\frac{|\alpha_H| \pi \ell^2 }{\beta})^{1/2}
\sum_k C_k (\pi^{1/2} L_y \ell)^{-1/2} \exp (i ky)
\exp (-(x-k\ell^2)^2/ 2 \ell^2)
\label{eq:2}
\end{equation}
where the number of terms in the sum over $k$ is
$ N_\phi = L_xL_y/2\pi \ell^2 $ and $ \ell^2 = \hbar /2eB $.
We define the following intensive thermodynamic
variables, which will be used to describe the system.
A dimensionless average local superfluid density is defined by
\begin{equation}
\Delta_0[C_k] = \frac{1}{N_\phi} (\frac{|\alpha_H| \pi \ell^2 }
                    {\beta})^{-1} \int d\vec r |\Psi (\vec r)|^2
         = \frac{1}{N_\phi} \sum_k |C_k|^2.
\label{eq:4}
\end{equation}
The participation ratio, defined by Eq.~(\ref{eq:a1}) is
\begin{equation}
P[C_k] = \frac{(\sum_k |C_k|^2)^2}
{\sum_{k_1k_2k_3k_4}
\bar C_{k_1} \bar C_{k_2} C_{k_3} C_{k_4}\delta_{k_1+k_2, k_3+k_4}
	\Theta(k_1,k_2,k_3,k_4)}
\label{eq:5}
\end{equation}
where $ \Theta (k_1,k_2,k_3,k_4) = (N_\phi L_x/Ly)^{1/2} \exp
            \lbrace -\frac{\ell^2}{2}
	[\sum_{i = 1}^{4}k_i^2-\frac{1}{4} (\sum_{i = 1}^4 k_i)^2] \rbrace$.
In terms of $\Delta_0 $ and $P$, the LLL-GL free energy has the following
form:
\begin{equation}
\frac{F_{GL}[C_k]}{N_\phi k_B T}  = g^2 (sgn (\alpha_H) \Delta_0[C_k] +
\frac{(\Delta_0[C_k])^2}{4 P[C_k]} ).
\label{eq:6}
\end{equation}
Temperature and field enter only through\cite{lllvalidity}
the dimensionless parameter,
$g \equiv \alpha_H (\pi \ell^2 / \beta k_B T)^{1/2} \propto
(T-T_{c2})/ (TH)^{1/2}$.

The partition function for the LLL-GL model is
\begin{equation}
	Z = (\frac{|\alpha_H|\pi \ell^2 }{\beta})^{N_\phi} \prod_k
           \int d\bar C_k d C_k \exp \{ -F_{GL} [C_k] /k_B T \},
\label{eq:7}
\end{equation}
and can be rewritten as the following form:~\cite{hu1}
\begin{equation}
	Z = \int d\Delta_0 d P \exp \{-N_\phi f(\Delta_0, P , g^2) \}
\label{eq:8}
\end{equation}
where
\begin{equation}
	f(\Delta_0, P, g^2) =  g^2 (sgn (\alpha_H) \Delta_0 +
	\frac{\Delta_0^2}{4 P} ) - s (\Delta_0, P)  - \ln ( |\alpha_H| \pi
	\ell^2 / \beta )
\label{eq:9}
\end{equation}
and
\begin{equation}
	s (\Delta_0, P  ) = \frac{1}{N_\phi}
	\ln \prod_k \int d\bar C_k dC_k
			\delta (P - P[C_k]) \delta (\Delta_0 - \Delta_0[C_k]).
\label{eq:10}
\end{equation}
Because $P [C_k] $ is invariant under a scale change
of $ |\Psi (\vec r) |^2 $, it follows that
\begin{equation}
	s (\Delta_0, P) = \ln \Delta_0 + s(P).
\label{eq:11}
\end{equation}
The participation-ratio entropy, $s (P) \equiv s(1, P)$,
expresses the portion of volume in
the phase space with a given participation ratio.

In the thermodynamic limit fluctuations in $\Delta_0[C_k]$ and $P[C_k]$ are
negligible so that the free energy of the system at a
given temperature ($g$) can be obtained by simply
minimizing $ f(\Delta_0, P , g^2) $ with respect to $\Delta_0$ and $P$.
It follows that the equilibrium values of $\Delta_0$ and $P$ satisfy
\begin{equation}
	g^2 (sgn (\alpha_H) + \frac{1}{2 P } \Delta_0) - \frac{1}{\Delta_0} = 0
\label{eq:12}
\end{equation}
and
\begin{equation}
	s^\prime (P) = -\frac{1}{4 P^2 }g^2\Delta_0^2.
\label{eq:13}
\end{equation}
Eq.~(\ref{eq:12}) establishes a functional
relationship between $\Delta_0$ and $P$
which originates in the fact that the properties of the system
do not depend on $\alpha_H$ and $\beta$ independently but only
on $g \propto \alpha_H/\beta^{1/2}$.  Note that for $g >> 1 $ \,
$\Delta_0 = g^{-2}$ while for $ g << 1$ $\Delta_0 = 2 P$.
Eq.~(\ref{eq:13}) then fixes the
equilibrium values of $P$ and $\Delta_0$ and hence the free energy.
This equation reflects the balance
between the rate of increase of
volume in order-parameter-space and the rate of decrease of condensation
energy which fixes the equilibrium participation ratio.
Note that the right hand side of Eq.~(\ref{eq:13}) vanishes for
$g >> 1$; the equilibrium value of $P$ in the high temperature limit
is the value of $P$ where $s(P)$ is maximized, {\it i.e.} the most
probable value of $P$ in order parameter space.
We remark a stable equilibrium can occur at any temperature
at participation-ratio $P$ only if
\begin{equation}
	s^{\prime\prime} (P) < 0.
\label{eq:14}
\end{equation}

We now turn to the evaluation of $s(P)$.  Since $P$ is positive
definite this function is defined on the interval
$(0, P_{\bigtriangleup})$ and we can evaluate it analytically
near both end points of the interval.  For $P$ near $P_{\bigtriangleup}$
we can approximate $P[C_k]$ by a Taylor series expansion to
second order around an Abrikosov vortex lattice state.
Since the Abrikosov lattice states occur at an extremum of $P[C_k]$
linear terms are absent and by formally diagonalizing
the resulting quadratic form we find that
the volume in order-parameter-space with a given value of $P$
is proportional to the surface area of a sphere in $\sim 2 N_{\phi}$
dimensions with radius proportional to $ (P_{\bigtriangleup}
- P)^{1/2}$.  It follows that for $P$ near $P_{\bigtriangleup}$
\begin{equation}
s (P) \approx \ln (P_{\bigtriangleup} - P ).
\label{eq:a2}
\end{equation}

The participation-ratio entropy for small $P$ is most easily evaluated
by using the symmetric gauge expansion of the order
parameter\cite{tesanovic1} in terms of eigenfunctions with
definite angular momentum, $m$.  In this case order parameters
which are confined to an area $\sim k/N_{\phi} A$ centered
on the origin, must be expanded in terms of the eigenfunctions with
$m < k$.  It follows that the order-parameter-space
volume with $P < k/N_{\phi}$ at $\Delta_0 = 1$ is given by the
surface area of a sphere in $k$ dimensions with radius $N_{\phi}^{1/2}$
and hence that for $P$ going to zero
\begin{equation}
s (P) \approx - P \ln P.
\label{eq:a3}
\end{equation}

We can also obtain analytic results for the expansion of $s (P)$
about its maximum by using the high-temperature expansion
of the free-energy of the GL-LLL model\cite{ruggeri1}:
\begin{equation}
F =  N_\phi k_B T ( \ln (\frac{\tilde \alpha_x}{\pi k_B T})
                         + f_{2D} (x))
\label{eq:19}
\end{equation}
where $\alpha_H = \tilde \alpha_x (1- 4 x)$,
$x \equiv (\beta k_B T)/(4\pi \ell^2  \tilde \alpha_x^2)$ and
$ g = (1-4x) / \sqrt{4x}$.
($x$ is non-negative,  $x = 1/4$ at $T_{c2}$,
$x \rightarrow  0$ for $ g \rightarrow + \infty$ and $x \rightarrow \infty$ for
$ g \rightarrow - \infty$.)
Coefficients in the power series expansion of $f_{2D}(x)$
may be evaluated using a diagrammatic perturbation expansion in terms
of self-consistent Hartree correlation functions.
The six leading coefficients were obtained by Ruggeri and Thouless
{}~\cite{ruggeri1} and the expansion
was later extended to eleventh order by Br\'{e}zin, Fujita and
Hikami~\cite{brezin2} and recently to thirteenth
order by us.~\cite{hu2}.
The equilibrium values of $\Delta_0$ and $P$ may be
expressed in terms of $f_{2D} (x)$ by differentiating the free energy
with respect to $\alpha_H$ and $\beta$.~\cite{hu2};
\begin{eqnarray}
P(x) & = & \frac{(1-2xf_{2D}^\prime (x))^2}
                {(4+(1-4x)f_{2D}^\prime (x))(1+4x)} \nonumber \\
    & = & \frac{1}{2} +  \frac{1}{2}\,x - \frac{8}{3}\,x^2 +
          \frac{452}{15}\,x^3
       - 431.59018759018753 \,x^4   + 7170.5968205856756 \,x^5 \nonumber \\
   &  & -134096.68933891651 \,x^6 + 2772357.9400791259 \,x^7
	- 62615750.777811569  \,x^8 \nonumber \\
   &  & + 1532019484.1067800 \,x^9  - 40349691260.478735  \,x^{10}
	+ 1138241888638.9989 \,x^{11}  \nonumber \\
    & & - 34247891779884.099  \,x^{12} + \cdots .
\label{eq:20}
\end{eqnarray}
Using Eq.~(\ref{eq:13}) and the high temperature expansion of
$\Delta_0 (x) $ we obtain the following expansion for
the parametric dependence of $s'(P)$ on $x$:
\begin{eqnarray}
s'(P(x)) & = &
 -\frac{(4+(1-4x)f_{2D}^\prime (x))^2\,x}{(1-2xf_{2D}^\prime (x))^2}  \nonumber
\\
 & = &  - 4\,x + 8\,{x}^{2} - \frac{260}{3}\,{x}^{3}
+ 1020.8000000000000\,{x}^{4} - 14012.376334776334\,{x}^{5} \nonumber \\
 &  & + 219842.58347280514\,{x}^{6} - 3870523.5408992074\,{x}^{7}
+ 75492832.182500102\,{x}^{8}   \nonumber \\
 & & -1615943686.2025482\,{x}^{9} + 37680981633.278216\,{x}^{10}
- 951344015946.40479\,{x}^{11} \nonumber \\
 & & + 25869507417290.368\,{x}^{12} + \cdots .
\label{eq:21}
\end{eqnarray}
Eq.~(\ref{eq:20}) can be inverted to expand $x$ in terms of $\tilde
P \equiv P-1/2$.  Inserting this series in Eq.~(\ref{eq:21}) gives
$s'(P)$ as a series in $\tilde P$.  Integrating this series and
determining $s(P=1/2)$ by direct evaluation we find
\begin{eqnarray}
S(P) & = & 1 + \ln \pi +  0 \,\tilde P -4\,\tilde P^2
- \frac{160}{9} \,\tilde P^3  + \frac{1096}{45} \,\tilde P^4
- 802.2613949013954 \,\tilde P^5  \nonumber \\
 & & + 16926.259949713371 \,\tilde P^6 - 499062.26189198034 \,\tilde P^7
+ 17336002.792339820 \,\tilde P^8 \nonumber \\
 & & - 688621672.74816198 \,\tilde P^9 + 30506759410.693500 \,\tilde P^{10}
- 1483164272081.5901  \,\tilde P^{11}  \nonumber \\
 & & + 78249334359919.989 \,\tilde P^{12}
- 4439326835886646.1 \,\tilde P^{13}
 + \cdots .
\label{eq:a4}
\end{eqnarray}

Directly evaluating the low-temperature expansion of the LLL-GL model
has proven to be an arduous task\cite{eilenberger1,thouless1,ruggeri2} and has
led to discordant results.  In contrast, using Eq.~(\ref{eq:a2})
we can obtain a result for the the leading low-temperature
correction to the mean-field free energy of the LLL-GL model in
a very simple way.  We find immediately, in agreement with
Ref.~\onlinecite{ruggeri2}, that for large $x$ ($ g << 1$ )
\begin{equation}
	f_{2D} (x) = -\frac{(1-4x)^2 P_{\bigtriangleup}}{4x} + \ln x.
\label{eq:22}
\end{equation}
The first term here is the mean-field-theory energy and the
correction comes from the low-temperature equilibrium
participation-ratio entropy.

It is interesting to observe that equilibrium values of
$P$ vary through a relatively narrow range between
the largest possible value for $P[\Psi]$
($P_{\bigtriangleup}$) and the most probable value for $P[\Psi]$ ($1/2$)
from low temperature to high temperature limits.  Useful
approximate expressions for the free energy, the magnetization, and
the specific heat of the GL-LLL system have been proposed
by Te\v{s}anovi\'{c} and collaborators\cite{tesanovic2,tesanovic4}
motivated by this property.

The high-temperature expansion of $s (P)$ around $P = 1/2 $ can be
extrapolated with the use of Pad\'{e} approximants for $s'(P)$.
The Pad\'{e} approximants were chosen to satisfy
$s'(P) = 1/(P-P_\bigtriangleup)$ for $P \to P_{\bigtriangleup}$.
and $s(P)$ was obtained by integration.
Results are shown in Fig.~\ref{fig:1}.  Poles appear in the
approximants for $\tilde P \sim -0.2$ and the extrapolation to
negative values of $\tilde P$ is not very successful.
Nevertheless, we believe that $s(P)$ is a smooth function
over the entire interval $(0,P_{\bigtriangleup})$.  This
expectation is consistent with numerical results for $s (P)$ obtained
by Monte Carlo methods which are also shown in Fig.~\ref{fig:1}.
The Monte Carlo results were extracted from distribution functions
for participation-ratio values, $ A_{\lambda} (P)$,  calculated using
$\exp (- N_{\phi} (\lambda P +\Delta_0)) $ as the sampling function.
Since $A_{\lambda} (P) \propto \exp ( N_{\phi} (s(P) - \lambda P) )$,
extrema of the distribution occur where $ s^\prime (P) =  \lambda $.
By performing calculations at a series of $\lambda$ values
we were able to map out the function
$s^{\prime} (P)$; the results shown in Fig.~\ref{fig:1} were obtained
by numerical quadrature from the Monte-Carlo results for $s^{\prime}
(P)$.  The overall agreement between the analytic and numerical results
for the participation-ratio entropy is excellent.  The inset in
Fig.~\ref{fig:1} shows Monte-Carlo results for $s^{\prime} (P)$ in
the narrow range of equilibrium participation-ratio values which
occur near the first-order melting transition.
We see that in the Monte-Carlo simulations (but not in the analytic
results) $s''(P) > 0$ for $P$ in the interval
$( 0.832, 0.837 )$.  Equilibrium $P$ values cannot occur in this
interval and must therefore have a discontinuity in their temperature
dependence.  It is this property of the participation-ratio entropy
which leads to the weakly first order phase transition in the LLL-GL
model.

Our description of the thermodynamics of the LLL-GL model
is summarized in Fig.~\ref{fig:2} in
terms of three contour plots for
$f(\Delta_0, P, g^2) + \ln (|\alpha_H|\pi\ell^2/\beta)$.
At each $g$ the equilibrium $(\Delta_0, P)$
minimizes $ f(\Delta_0, P, g^2) $.
The top panel is for a temperature at which the system
is in the vortex liquid state, the middle panel
is for a temperature close to the phase transition and the bottom
panel is for a temperature at which the system is in the vortex
lattice state.  The trend to decreasing participation ratios at
higher temperatures is driven by the increase in the
relative importance of the entropy.  The first order phase
transition occurs because of the occurrence of an interval
over which $s''(P)$ is positive.  When the optimal $P$ values
are close to this interval two local minima appear in the
free energy contours and the global minimum switches from the
local minimum at larger $P$ to the local minimum at smaller $P$
as the temperature increases.  In the high temperature limit
of the model, thermal fluctuations are Gaussian and $P$ approaches $1/2$.

This work was supported by the Midwest
Superconductivity Consortium through D.O.E.
grant no. DE-FG-02-90ER45427.   The authors are grateful to
Steve Girvin and Zlatko Te\v sanovic' for helpful conversations.

\begin{figure}
\caption{Analytic and numerical data for $s (P) $.
The inset shows Monte Carlo results for $s' (P) $. }
\label{fig:1}
\end{figure}

\begin{figure}
\caption{
Contour plots for
$f(\Delta_0, P, g^2) + \ln (|\alpha_H|\pi\ell^2/\beta) $ at
$g = -5.5$(top panel, vortex liquid state), $g = -6.6$ (middle panel, at
phase transition ) and $g= -7.1$  (bottom panel, vortex solid state).
}
\label{fig:2}
\end{figure}


\begin{thebibliography}{10}
\bibitem{lllvalidity}
  The applicability of the two-dimensional LLL-GL model
  to experimental systems has been established by verifying that temperature
  and magnetic field enter only in the combination $(T-T_{c2})/(TH)^{1/2}$
in Ref. \onlinecite{urbach1}.
\end{thebibliography}
\end{document}